\documentclass[a4paper,11pt]{article}
\pdfoutput=1 % if your are submitting a pdflatex (i.e. if you have
\usepackage{jheppub} % for details on the use of the package, please
\usepackage[T1]{fontenc} % if needed

\def\bt{\bar{t}}

\def\bx{\bar{x}}

\newcommand{\be}{\begin{equation}}
\newcommand{\ee}{\end{equation}}
\newcommand{\ba}{\begin{array}}
\newcommand{\ea}{\end{array}}
\newcommand{\bea}{\begin{eqnarray}}
\newcommand{\eea}{\end{eqnarray}}
\newcommand{\mat}[4]{\left(\begin{array}{cc} #1 & #2 \\
    #3 & #4 \end{array}\right)}
\title{\boldmath Calabi's diastasis as interface entropy}

\author[a,b]{Constantin P. Bachas,}
\author[b,c]{Ilka Brunner,}
\author[d,e]{Michael R. Douglas,}
\author[ e, f]{Leonardo Rastelli\,}
 
\affiliation[a]{  Laboratoire de Physique Th\'eorique de l'\'Ecole Normale Sup\'erieure,$^1$ \note{Unit\'e mixte 
(UMR 8549) du CNRS 
et de l'ENS, Paris.
}\\
 24 rue Lhomond, 75231 Paris cedex, France}  
\affiliation[b]{Arnold Sommerfeld Center, Ludwig-Maximilians-Universit\"{a}t,\\
Theresienstr. 37, M\"{u}nchen 80333, Germany}
\affiliation[c]{Excellence Cluster Universe, Technische Universit\"{a}t M\"{u}nchen,\\
Boltzmannstr. 2, Garching 85748, Germany}
\affiliation[d]{Simons Center for Geometry and Physics, 
Stony Brook University, \\ Stony Brook, NY 11794,  USA}
\affiliation[e]{C.N. Yang Institute for Theoretical Physics,
Stony Brook University,\\
Stony Brook, NY 11794-3840,  USA}
\affiliation[f]{Institute for Advanced Study, Einstein Dr., Princeton, NJ 08540, USA}

\emailAdd{bachas@lpt.ens.fr}
\emailAdd{ilka.brunner@physik.uni-muenchen.de}
\emailAdd{mdouglas@scgp.stonybrook.edu}
\emailAdd{leonardo.rastelli@stonybrook.edu}

\preprint{\\ YITP-SB-13-37\\  LMU-ASC 77/13\\ LPTENS-13/21}
\bigskip

\abstract{We  show that the entropy  of certain conformal interfaces between  
 $N=(2,2)$ sigma models that belong to the same moduli space, 
 has a natural geometric interpretation in the large volume limit as  Calabi's diastasis function. 
 This is an extension of the well-known relation between the quantum K\"ahler potential
 and the overlap of canonical Ramond-Ramond ground states in $N=(2,2)$ models.
  }

\begin{document} 
\maketitle
\flushbottom

\section{Introduction}

 The purpose of the present note is to establish a geometric formula for the entropy   of certain
 superconformal interfaces between $N=(2,2)$ superconformal  sigma models. 
   As  is well known, in the large volume limit the target spaces of such sigma models are
  Calabi-Yau manifolds.  
   The interfaces of interest separate theories with the same K\"ahler modulus
  but different complex structure, or vice versa, and they reduce to the trivial interface when the moduli
  of the two theories coincide. 
  Our main result  is that for such interfaces
  \bea\label{calabi-dist}
  2\, {\rm log}\, g =  K(t ,\bar t ) + K(t^\prime,\bar t^\prime) - K(t ,\bar t^\prime) - K(t^\prime,\bar t ) ,
  \eea
 where $g$ is the universal degeneracy  \cite{Affleck:1991tk} of the interface \footnote{When the interface is viewed as an operator between the initial and deformed theory, then $g$ is the image of the identity projected to the identity of the other theory, see Section 4 for more details.} , 
   $t$ and $t^\prime$ are the moduli of the theories on either side of the interface, and $K$ is the K\"ahler
 function on moduli space. 
 
 \smallskip 
  The right-hand side of  the above equation is a known quantity in K\"ahler geometry; it is the so-called
{Calabi diastatic function} \cite{Calabi}.
 It can be defined on any K\"ahler manifold (this requires showing that the analytic continuation
of $K(z,\bar z)$ to independent $z$ and $\bar z$ makes sense, which is done in \cite{Calabi}).
A nice feature of the combination \eqref{calabi-dist} is that the K\"ahler-Weyl dependence of $K(z,\bar z)$
cancels out, so that it is a function.  It agrees with the geodesic distance at small separations, but
has the property that it is preserved under restriction to a submanifold.

\smallskip
  Equation \eqref{calabi-dist} gives a world-sheet definition of the diastatic function that can
  be used away from the geometric, large volume limit. 
  It  is a natural extension of the  well-known formula that relates the (quantum) K\"ahler potential to
   the norm of a canonical Ramond-Ramond (RR) ground state in  the sigma model, 
     \cite{Periwal:1989mx,Cecotti:1991me}   
  \bea
   \ _{\rm\footnotesize  RR}\langle   \bar 0    \vert 0 \rangle_{\rm\footnotesize  RR} 
 \ =\   e^{ - K(t, \bar t) }\ . 
\eea 
  Recently,  \cite{Jockers:2012dk,Gomis:2012wy} 
  the norm of this RR ground state has been related to the  partition function of $N=(2,2)$  gauge
  theories on the (squashed) two-sphere,  which can be computed  exactly
  using the technique of localization  \cite{Benini:2012ui,Doroud:2012xw}. This is a new way to
  compute world-sheet instanton corrections to the K\"ahler potential, and to extract  Gromov-Witten invariants, 
  without 
  the need to identify and   solve  a classical geometric  mirror problem.    
 
 Similarly, through equation \eqref{calabi-dist} one may relate  quantum  corrections  to Calabi's diastasis function to
    the partition function on the (squashed) two-sphere in the presence
 of certain $N=2$ supersymmetric domain walls.  Localization techniques for the computation of  the latter have been
 developed recently in
  \cite{Hori:2013ika,Sugishita:2013jca}. They could be used to extract the relevant  ``open-string''  Gromov-Witten invariants,
  which are  notoriously hard  to compute   by  other means.
\smallskip

      We were actually led to this formula while studying the following broader question:   how to 
     define  alternative metric(s) on
      spaces of  conformal field theories \cite{Douglas:2010ic}\,?  One promising proposal  \cite{inprogress} is to define 
      the  distance between CFT$_1$ and CFT$_2$ as
      \bea\label{distance}
      d(1,2) = {\rm min}_{\, \cal S}\, \sqrt{\log\, g}\ , 
      \eea
      where ${\cal S}$ is  an
       appropriate set of   interfaces which separate the two conformal theories.
   An appealing feature of such a definition is that $d$  reduces to the Zamolodchikov metric whenever
   CFT$_1$ can be obtained from  CFT$_2$  by a small variation of continuous  
   moduli. 
    In the special case of $N=(2,2)$ models this follows immediately from \eqref{calabi-dist},
   but the proof is more general  \cite{inprogress} and does not require the use of supersymmetry.  
Another appealing feature of the above definition  is that CFT$_1$ and CFT$_2$ need not  belong
to the same moduli space, or even have the same central charge. In particular, equation \eqref{calabi-dist}  
extends the definition of the diastatic  function to  pairs of sigma models  separated by an $N=2$ interface,
even when these sigma models belong to 
 different moduli spaces. 
 \smallskip

  Despite its intuitive appeal, the  proposal \eqref{distance} does not automatically obey the axioms for a proper distance.
  In particular, conformal  interfaces may have negative entropy, and
  Calabi's  diastasis need not always obey, as we will show,  the triangle inequality. 
 Ideas for  bypassing these obstructions, by 
 restricting the set  $\cal S$ of allowed interfaces,  will be discussed elsewhere  \cite{inprogress}.
  Here we concentrate on proving  formula  \eqref{calabi-dist},  which  is 
  interesting in its own right as a new  entry in the   ``worldsheet versus target-space geometry''
 dictionary.
 \smallskip
 
  The paper is organized as follows: in section~\ref{sec2} we prove   formula
  \eqref{calabi-dist}
   in the simplest case of $N=(2,2)$
  sigma model whose target space is the two-dimensional torus. We give both an algebraic and a geometric  
  derivation of $g$ for any  moduli deformations, and show that it reduces to \eqref{calabi-dist}   when either the K\"ahler or
  the complex structure are held fixed. 
  In section~\ref{sec3} we extend the geometric derivation  to arbitrary  Calabi-Yau $n$-folds with $n>1$. 
  This uses the well-known ``folding trick'', to map the interfaces to branes in a product Calabi-Yau manifold.    
  Section~\ref{sec4}  presents  an   algebraic derivation of this result, which only relies on the $N=2$ supersymmetry  of the interface. 
  This  shows that interface entropy  provides  a natural extension of Calabi's diastasis  in the non-geometric regime, 
  and  even when  the two
  worldsheet theories  do not belong to the same moduli space. 
    Finally, in  section ~\ref{sec5} we show that Calabi's  diastasis function does not    obey
      the triangle inequality in spaces with positive sectional curvature, and may hence 
  fail   one of the key tests for a proper distance. We  
  conclude the section with some  remarks.

    %%%%%%%%%%%%%%%%%%%%%%%%%%%%  
   
\section{The two-dimensional torus CFT}
\label{sec2}

 The simplest Calabi-Yau manifold   is the two-dimensional torus,  $T^2= \mathbb{R}^2/(\mathbb{Z}\times \mathbb{Z})$.  
 As a warm up, we shall  first  derive the  
formula \eqref{calabi-dist} in this  special  case. 
\smallskip

We parametrize the torus by $(x , y) \in (0,1]\times (0,1]$. 
 The  K\"ahler and  complex structure moduli,  $\tau = \tau_1 + i \tau_2$ and $\rho = \rho_1 + i \rho_2$,  are 
  related   to the flat  metric, $G$,  and antisymmetric Neveu-Schwarz field,  $B$,  as follows:\footnote{Later, we
  will refer to the complex structure moduli collectively as $t$, and to the K\"ahler moduli as $u$. But for
   the 2-torus we use  the
  more canonical notation,  $\rho$ and $\tau$.}
  \be\label{2torusModuli}
 G = {  \tau_2\over  \rho_2}  
 \left(\begin{array}{cc}   1 &     \rho_1  \\     \rho_1 &     \vert \rho\vert^2   \end{array}\right) \ \ , \qquad
 B =  \left(\begin{array}{cc}   0  &   \tau_1   \\     -\tau_1 &    0   \end{array}\right) \ .  
\ee
In terms of  the complex coordinate $z = x  + \rho y$ one has
\be
ds^2 = {  \tau_2\over  \rho_2}\, dz d\bar z \
\Longleftrightarrow\
  k  = {i\over 2} {\tau_2\over \rho_2}\, dz\wedge d\bar z = \tau_2\, dx \wedge dy\ ,
\ee 
where $k$ is the real K\"ahler form. It is   complexified by the addition of the Neveu-Schwarz 2-form,  
$\omega = B +  i k\  $ with 
  $\ B   =   \tau_1\, dx \wedge dy $. 
The holomorphic $(1,0)$ form is $\Omega = dz$, up to an irrelevant  multiplicative constant. 
\smallskip

  The moduli space of  $N=(2,2)$ superconformal theories
    with     target space  $T^2$  consists of two copies of  the symmetric coset
  ${\cal M} = SL(2, \mathbb{Z}) \backslash SL(2, \mathbb{R})/ SO(2)$. One copy parametrizes the complex structure  modulus,  
  and the other  the K\"ahler modulus. 
 T-duality  exchanges $\tau$ and $\rho$, so that the full moduli space is 
 $({\cal M}\times {\cal M})/\mathbb{Z}_2$. 
  The  metric on this  moduli space
  derives  from the K\"ahler potential
  \be
  K = K_{\rm K} (\tau, \bar\tau) + K_{\rm C} (\rho, \bar\rho)\ , 
  \ee
 where $K_{\rm K}$ and $K_{\rm C}$ are given by  
 \be\label{KahlerTorus}
 K_{\rm K} = - {\rm log}   \left(   \int_{M}  k  \right)  =  - {\rm log}\, \tau_2 \ , \qquad
 K_{\rm C} = -  {\rm log}\left( \int_{M} {i\over 2}\, \Omega \wedge \bar\Omega\right) = - {\rm log}\, \rho_2\ . 
 \ee 
   The $SL(2, \mathbb{R})$   transformations of $\tau$ and $\rho$  
  act  as   K\"ahler-Weyl
  transformations on the K\"ahler potential,  $K\to K + f + \bar f$ where $f$ is  a holomorphic function
  of $\tau$ and $\rho$. Such transformations leave the metric  invariant,  as expected.

%%%%%%%%%%%  
  
\subsection{Algebraic derivation}

  Consider now two conformal theories with moduli $(\tau, \rho)$ and   $(\tau^\prime, \rho^\prime)$.
  We will be  interested in a special conformal interface between these two  theories -- the ``deformed identity''
  interface  introduced and discussed in \cite{Bachas:2001vj,Bachas:2007td,Brunner:2008fa, Bachas:2012bj}. 
  This is the deformation 
   of  the  trivial defect (the  ``no  interface'')  as the moduli of the second CFT   vary  continuously 
    from $(\tau, \rho)$ to   $(\tau^\prime, \rho^\prime)$.
         
      In general, such an interface could depend on the specific deformation path,
    as well as on  (open-string) moduli. 
      However, in the case at hand,  the deformed identity only depends on the homotopy class of the
      deformation path, and its 
    $g$-function is independent  of open-string moduli. Thus  log$\,g$ is
     a well-defined
   function of  pairs  of points on the covering space  of moduli space, i.e. on 
   two copies of the upper-half complex plane.

    \smallskip   
     Both the trivial interface,  and its deformations,  preserve the $U(1)^4$ symmetry of the toroidal theory. 
   Such symmetry-preserving  interfaces  were analyzed recently
   in   \cite{Bachas:2012bj}, where it was shown that their
     $g$-function  can be written as
   \be\label{BBR}
   {\rm log}\,  g =  {1\over 2} {\rm log\, det} (\Lambda_{22})\ , \qquad {\rm with} \qquad
   \Lambda =  {\footnotesize \mat{ \Lambda_{11}}{\Lambda_{12}} {\Lambda_{21}}{\Lambda_{22}} }
   \ee
    the $SO(2,2)$ matrix that relates the  even self-dual Lorentzian charge lattices of the two theories. 
   Written 
     in $2\times 2$-block form,  this matrix  obeys $\ \Lambda^t {\scriptsize \mat{0}{\bf 1}{\bf 1}{0} \Lambda} = {\scriptsize \mat{0}{\bf 1}{\bf 1}{0}  }$.

  One can give an explicit formula for the matrix $\Lambda$ corresponding to  the deformed
%%%%%%
   identity
  by using the expression for the charge lattice 
   of toroidal models  
   in terms of   the metric $G$ and Kalb-Ramond field $B$  \cite{Giveon:1994fu}.   The answer  is\,\footnote{More general
  $U(1)^4$ symmetric    interfaces  are given by  $ \Lambda = V^\prime \hat\Lambda V^{-1}$  where $\hat\Lambda$ is an element of $O(2,2, \mathbb{Q})$.
  These  have  $g= \sqrt{l.c.m.\times \vert\Lambda_{22}\vert}$ where $l.c.m.$ is the least common multiple of the matrix elements of $\hat\Lambda$
  \cite{Bachas:2012bj}.
  }
 %%%%
   \be
   \Lambda = V^\prime V^{-1} \   \qquad {\rm with}\qquad V = \mat{\hat e}{\ \hat e (B+G)}{\hat e}{\ \hat e(B-G)}\ ,  
   \ee
  where $\hat e$ is the vielbein that satisfies  $2\, \hat e^t\, \hat e  = G^{-1}$, and there is a similar
  expression for $V^\prime$. Inserting these formulae and \eqref{2torusModuli}
   in equation \eqref{BBR} leads to the following expression for
    the  $g$-function of the deformed identity:   
  \be\label{torusg}
 g_{\rm d.i.}  =   \left[  { (\tau - \bar\tau^\prime)(\tau^\prime - \bar\tau) \over (\tau - \bar\tau)(\tau^\prime - \bar\tau^\prime)} 
 +  { (\rho - \bar\rho^{\,\prime})(\rho^\prime - \bar\rho) \over (\rho - \bar\rho)(\rho^\prime - \bar\rho^{\,\prime})}  - 1 \right]^{1/2} \ . 
\ee
 As  anticipated earlier, this is a well-defined function of $\tau^\prime, \rho^\prime$ as these range  over
  the (simply-connected) covering space of the CFT moduli space.
Note that $g_{\rm d.i.}$   is   invariant under a simultaneous $SL(2,\mathbb{Z})$  transformation  of  the primed and unprimed moduli, 
     but not  under a transformation of   only one  of the two CFTs. 
  
%%%%%%%%%%%%%%%%  

\subsection{Geometric derivation}

    The expression \eqref{torusg} can be derived more directly, in a way that will
generalize  to any large volume 
     Calabi-Yau $n$-fold.
  The starting point    is the folding trick, which maps an  interface  between
   $\sigma$-models  with target spaces $M$ and $M^\prime$ to a boundary of the $\sigma$-model
    with target space $ M \times M^\prime$ \cite{Bachas:2001vj,Bachas:2004sy,Fuchs:2007fw}.   
 When the   two $\sigma$-models are identical  there exists a trivial interface across which all the fields are continuous --
 the ``no interface''.  This is
 mapped after folding to the  diagonally-embedded middle-dimensional D-brane:     
   $M\rightarrow M\times M$ given by $x\rightarrow (x,x)$.  Now as one of the   $\sigma$-models is deformed, 
 this diagonal  brane is also  deformed to a  new brane, $\Delta_f$, which we describe (at least locally)  as the graph of a function $f$
 from $M$ to $M^\prime$. 
 Put differently,   $\Delta_f$ is given by the embedding  $x\rightarrow (x, f(x))\in M \times M^\prime$. 
 \smallskip 
    
      We may determine   $f$ by minimizing the $g$-function of the brane --  this is the condition of conformal invariance.  
    In the large volume limit, the $g$-function is the appropriately normalized  Dirac-Born-Infeld action
        (see e.g.  \cite{Harvey:1999gq}): 
\be\label{DBI}
g \simeq  \frac{ \int_{M} {\rm det}^{1/2}( G- B + f^*G^\prime + f^*B^\prime) }
{\sqrt{2^d\, \ Vol \hskip -1mm (M) \ Vol \hskip -1mm (M^\prime)}}\ . 
\ee
Here $f^*$ denotes the pullback from $M^\prime$ to $M$, and we note
  that folding  
      flips the sign of  $B$  thereby complex-conjugating the K\"ahler form of the folded $\sigma$-model. 
   In the toroidal case at hand this  amounts to trading  $(\tau, \rho)$ for  $(-\bar\tau,  \rho)$.  
  Note also the  normalization of the DBI action by the volume factors in \eqref{DBI}. This 
     can be  fixed by requiring that   $g =1$ for the trivial (or identity)   interface.\,\footnote{In string theory, the $g$-function
 of a D-brane wrapping some dimensions of  the compact space is the mass of the corresponding point-particle 
  in  the Einstein frame.}  

\smallskip

  For  toroidal theories,  the  identity defect is a diagonally-embedded planar D2-brane. As  one
   of the two tori is deformed this D2-brane follows suit, i.e. it is still given by 
  the planar diagonal embedding $x=x^\prime$ and $y=y^\prime$ where $(x,y)$ and $(x^\prime, y^\prime)$ are the canonically-normalized
  flat coordinates of the two tori. 
One thus  finds
 \be
 g_{\rm d.i.} =  {1\over \sqrt{4\tau_2 \tau_2^\prime}} \,  {\rm det}^{1/2}  ( G + G^\prime -  B +  B^\prime ) \ ,  
\ee
which after  inserting  (\ref{2torusModuli})  and doing some simple  algebra leads to  the result \eqref{torusg}
for the $g$-function of the deformed identity. 
 Note that the two, geometric and algebraic,  derivations of $g$ give the same result, because
 the DBI approximation  \eqref{DBI}  is in this case exact. 

\smallskip

%%%%%%%%%%%%%

\subsection{Supersymmetry  and diastasis}
  
Expression \eqref{torusg} simplifies considerably if $\rho = \rho^\prime$, i.e. if one keeps the complex structure  fixed and only deforms
the K\"ahler modulus of the torus. The folded interface is in this case   the holomorphic brane $z= z^\prime$, and 
 \be
 2\,  {\rm log} \, g_{\rm d.i.}\Bigl\vert_{\rho=\rho^\prime}  \ =   - {\rm log} (\tau - \bar\tau) -  {\rm log} (\tau^\prime - \bar\tau^\prime) 
 +  {\rm log} (\tau - \bar\tau^\prime) +  {\rm log} (\tau^\prime - \bar\tau) \ , 
\ee
which is precisely Calabi's diastasis  function for the  potential  \eqref{KahlerTorus}.   

\smallskip
 
The same conclusion  holds  if one only deforms   the complex structure, $\rho$,  keeping the  K\"ahler modulus, $\tau$,   fixed.
The $g$ function of such branes is given again  by Calabi's diastasis,  
 \be
 2\,  {\rm log} \, g_{\rm d.i.}\Bigl\vert_{\tau=\tau^\prime} \  =   - {\rm log} (\rho - \bar\rho) -  {\rm log} (\rho^\prime - \bar\rho^\prime) 
 +  {\rm log} (\rho - \bar\rho^\prime) +  {\rm log} (\rho^\prime - \bar\rho) \ .  
\ee
   This  is  of course expected by mirror symmetry. For later use,  it is
   nevertheless  interesting to understand how supersymmetry is preserved
    in this case. 
  
  To this end, we consider the 2-form  $\bar\Omega\wedge \Omega^\prime = d\bar z\wedge dz^\prime$. 
The  D-brane corresponding to the folded interface  obeys trivially
\be\label{11}
 {\rm Im} (e^{i\theta}\, d\bar z\wedge dz^\prime)  \Bigl\vert_{\Delta_f} =  {\rm Im} [e^{i\theta}\,  (\rho^\prime - \bar\rho) dx\wedge dy]
 \Bigl\vert_{\Delta_f}  = 0 \ , 
\ee
where $-\theta$ is the  phase of the complex number $( \rho^{\, \prime} -  \bar\rho)$. 
Furthermore, since $\tau = \tau^\prime$,  the restriction of the two K\"ahler forms on the D-brane is the same,   
\be
(k- k^\prime) \Bigl\vert_{\Delta_f} = 0\ .  
\ee
Finally the following two top-forms are equal to the volume form of the doubled torus,
  up to an irrelevant multiplicative constant:  
\be\label{13}
(d\bar z\wedge dz^\prime)\wedge (d  z\wedge d\bar z^{\, \prime}) = {\cal C} (k-k^\prime)^2\ . 
\ee
The set of conditions  \eqref{11}-\eqref{13} define special lagrangian submanifolds,  
 which  preserve   $N=2$ supersymmetries in any Calabi-Yau space. 
 \smallskip

 In conclusion, interfaces between  theories which differ  only in complex structure, 
 or in K\"ahler form, preserve half of the bulk supersymmetries, and their entropy is the  diastasis function. 
    Note  that if  one  varies both $\tau$ and $\rho$,  
 then  \eqref{torusg}  is not  any more related to the  diastasis function.\,\footnote{The reader may here object that
 any  planar brane in a four-torus is half-BPS, and this would continue to be true for the direct product of any two tori. 
 The unbroken supersymmetries mix however in this case the 
 fields of the two tensored CFTs, so they  are
 not  local symmetries after unfolding.}
    Nevertheless, these  two functions  do coincide  for  small  deformations at quadratic order. 
   This is actually a  general fact:
  $\sqrt{{\rm log} \, g }$ of the deformed-identity  can be shown to reduce  to the Zamolodchikov distance
   for {\it all}  infinitesimal marginal deformations of a 2d conformal theory, 
   whether they preserve supersymmetry or not  \cite{inprogress}.

 %%%%%%%%%%%%%%%%%%%%%%%%%%%%  
   
  \section{Large volume Calabi-Yau $\sigma$-models}
  \label{sec3}
  
   It is straightforward to extend the  geometric arguments of the previous subsection to any
   Calabi-Yau sigma model  in the large volume   limit. 
The product  of two Calabi-Yau $n$-folds, $M\times M^\prime$,  
 is also a Calabi-Yau manifold of complex dimension $2n$. 
Its K\"ahler form is $k+k^\prime$, and its holomorphic
$(2n,0)$ form  $\Omega \wedge \Omega^\prime$.  
Like all  Calabi-Yau manifolds, $M\times M^\prime$ has two types of supersymmetric submanifolds  
  \cite{ Becker:1995kb,Bershadsky:1995qy,Ooguri:1996ck,Brunner:1999jq} :  
  the special Lagrangians (A-type),  and the
holomorphic submanifolds (B-type).  As we will see, these correspond to interfaces between theories
with the same complex, respectively K\"ahler,  structures. 
 
\subsection{K\"ahler structure deformation}

  Consider  the trivial interface between two identical $\sigma$-models, which after folding becomes the diagonal brane
$M\rightarrow M\times M$ given by $x\rightarrow (x,x)$. This is a holomorphic brane since in complex coordinates
 we can write $z\rightarrow (z,z)$. Let us now deform one of the theories from $M$ to $M^\prime$. 
  If $M$ and $M^\prime$ have the same complex structure,  the  above 
  brane will stay  holomorphic and, in general, 
  there will be no other nearby holomorphic branes. The $g$-function of this deformed identity is
   proportional,  in the large volume limit, 
  to the Dirac-Born-Infeld action 
   \be\label{31}
g_{\scriptsize\, \rm hol} \  \simeq \  { 2^{-n} \vert  \int_M (-\bar \omega +\omega^\prime)^{n}  \vert \over  \vert \int_M k^{\,n} \vert^{1/2}\,
  \vert \int_M (k^{\prime})^n   \vert^{1/2} } \ ,
\ee
  where $\omega$ and $\omega^\prime$ are the complexified K\"ahler forms of $M$ and $M^\prime$, and we recall 
   that folding transforms $\omega := B + ik$  to $-\bar \omega$. As explained in the previous section,   the 
    normalization factor  can be fixed by requiring   that 
 $g_{\scriptsize\, \rm hol} = 1$ for the identity, namely  when  $\omega=\omega^\prime$.

 Taking the logarithm of \eqref{31},  and using the fact
 that $K (u, \bar u) \simeq  - {\rm log}(\int_M k^n )$  gives precisely Calabi's diastasis
   \be  \label{eq:cy-distA}
2 \, {\rm log}\, (g_{\scriptsize\, \rm hol}  )   \simeq  K(u,\bar u ) + K(u^\prime,\bar u^{\, \prime}) - K(u ,\bar u^{\, \prime})  -  K(u^\prime,\bar u ) \ , 
\ee 
 where $u $ denotes collectively  the K\"ahler moduli, and we have defined  the 
   analytic extension of $K(u,\bar u )$  to independent   $u$ and $\bar u$   as follows
   \be \label{eq:cpx-kpot}
K(u^\prime ,\bar u)\simeq - \log   \int_M [  -\bar \omega(\bar u)  + \omega(u^\prime) ]^n  + n \log\, 2 \ . 
\ee
 We expect this last formula to make sense in any open neighborhood of a generic point on the K\"ahler cone. 
   That  the  analytic continuation of  $K(u, \bar u)$ makes sense (and is unique)
    is  a crucial  input   in the original work of Calabi. 
  \smallskip
 
 This proves the equation \eqref{calabi-dist} for interfaces separating
  theories with the same complex structure but different K\"ahler
 forms, in the large volume limit.

 %%%%%%%%%%%%%%%

 %%%%%%%%%%%%%%%
 
 \subsection{Complex structure deformation}
 
    What is the  the mirror statement to  \eqref{eq:cy-distA}?  If $M$ and $M^\prime$ have the same K\"ahler
    form but different complex structures,
    then the deformed identity interface does not correspond to  a holomorphic submanifold. 
  We will now show  that it corresponds to a special Lagrangian
 (sLag) brane calibrated not by the holomorphic $(2n,0)$ form, but  by the appropriately-normalized  
 mixed $2n$-form  $\varphi =  i^n\,  \bar \Omega \wedge \Omega^\prime$.  
\smallskip

 The existence of this extra calibrating form, in addition to the holomorphic volume form, is due to the fact that 
$M\times M^\prime$   is ``more special'' as it has $SU(n)^2\subset SU(2n)$ holonomy. 
 One may in particular   complex conjugate one of the two manifolds in the product, which
 gives also  a new symplectic form $s=  i(k - k^\prime)$, different from  the standard K\"ahler form $k+k^\prime$. 
 Clearly, both $\varphi$ and $s$ are closed forms, and they satisfy, up to normalization, 
the top-form condition
\be
2^{-2n} \varphi \wedge\bar  \varphi = {1\over (2n)!} \, s^{2n} = d(volume) \ .  
\ee
They can  thus be used to construct  a new class of sLag submanifolds,\,\footnote{This observation has been  
exploited, for instance,   in the context of $N=(2,0)$ $\sigma$-models in  \cite{Kapustin:2010zc}.}
different from  those constructed with the standard K\"ahler and holomorphic volume forms. 
\smallskip

  It is  easy to see that the  brane $\Delta_{id}$  corresponding  to the trivial interface belongs to this new class. 
  The   necessary and sufficient  conditions  (see e.g. \cite{Joyce:2001xt}),
  \be
   s \Bigl\vert_{\Delta_{id}} = 0 \qquad {\rm and} \qquad  {\rm Im} \varphi\Bigl\vert_{\Delta_{id}} = 0\  ,
  \ee
  are  trivialy satisfied in this case.  The sLag conditions can  still be imposed if we deform the complex
  structure of $M^\prime$ without changing its K\"ahler form (so that $k= k^\prime$).
  The Lagrangian requirement for the submanifold $\Delta_f \subset M\times M^\prime$ now reads
  \be
   k = f^*k \ , 
  \ee
which  says that  $f$ preserves   the  K\"ahler form --  it is    a ``symplectomorphism''. 
Many such maps indeed exist, and  they can be specified locally by a single function  $F({\rm Re}z, {\rm Re}z^\prime)$ 
which defines a canonical transformation. This function can then be determined by the 
(volume-minimizing) calibration  condition
\be\label{calibr}
{\rm Im} (e^{i\theta} \, \varphi)\Bigl\vert_{\Delta_{f}}  =   {\rm Im} (e^{i\theta} i^n  \, \bar\Omega\wedge f^* \Omega^\prime ) = 0\,    
\ee
  for some constant phase $\theta$. This gives an equation for the  function $f$,  which can be always solved  at least in a local patch. 
  In the toroidal theory of the previous section, $f$ is the map $(x^\prime, y^\prime) = (x,y)$ as the reader can easily verify. 
 
 \smallskip 
  
  It has been actually shown \cite{McL}  that for compact manifolds   the moduli space of a sLag submanifold has dimension
      equal to its  first Betti number.  By continuity this should be  in our case  the first Betti number of  $\Delta_{id}$,  which is isomorphic to 
      the Calabi-Yau manifold $M$. Since $b_1(M)$  vanishes  for complex dimension $n>1$, the  sLag submanifold   $\ \Delta_{f}$ is unique 
       in all cases with the exception of   the two-torus.\footnote{For the 2-torus  the sLag D2-brane has $b_1=2$. Its 
        two geometric moduli,  which determine the brane's position on the doubled torus, 
      get  complexified  by  the  Wilson lines along the two non-contractible cycles.}
             
  \smallskip     
       
               The $g$-function of this sLag D-brane is given by its volume,  which by the sLag condition 
               is the integral of the calibrating form 
   \be\label{sLagg}
 g_{\scriptsize \, \rm sLag}=  { \vert  \int_M \bar \Omega \wedge f^*\Omega^\prime \vert   \over 
( \vert  \int_M   \Omega \wedge \bar  \Omega\vert\,  \vert   \int_M   \Omega^\prime  \wedge \bar  \Omega^\prime \vert )^{1/2} }\ . 
   \ee
 The normalization was once again fixed so as to ensure that $g=1$ for the trivial interface. Notice that we do not need here  the
 Dirac-Born-Infeld action, because the $B$-field  on the  D-brane is zero. This is because we have assumed that
 the two $\sigma$-models  have the same  {complexified} K\"ahler moduli, i.e.  $k= k^\prime$  {and}  $B= B^\prime$. 
  Since folding flips the sign of $B$ in one of the models,  the net field on the sLag brane vanishes. 
  \smallskip  
  
 Expression \eqref{sLagg}  is again suggestive of an exponentiated diastasis function for the K\"ahler potential
 in complex structure moduli space. We denote complex moduli by $t$. 
 % . 
 The analytic extension of  $K(t,\bar t ) = -\log  \int_M   \Omega(t)\wedge\bar\Omega(\bar t)$
 suggested by \eqref{sLagg} is
  \be \label{eq:cpx-kpot}
K( t^\prime, \bar t) = -\log  \int_M  f^*\Omega(t^\prime)\wedge\bar\Omega(\bar t) \ ,  
\ee
with $f$ defined by the calibration condition \eqref{calibr}. Notice that this condition only depends on $t^\prime$ and $\bar t$,
so the function $f$ does not introduce any implicit dependence on the conjugate variables,  $\bar t^\prime$ and $t$,
as claimed.\,\footnote{One
 may of course  complex-conjugate \eqref{calibr} and express the calibrating map $f$, equivalently,  in terms only of the independent
variables $\bar t^\prime$ and $t$.}
With this analytic extension one has
   \be  \label{eq:cy-distB}
2 \, {\rm log}\, (g_{\scriptsize\, \rm sLag}  )  =   K(t,\bar t ) + K(t^\prime,\bar t^{\, \prime}) - K(t ,\bar t^{\, \prime})  -  K(t^\prime,\bar t ) \ , 
\ee 
which proves the advertised identity  \eqref{calabi-dist} for interfaces between
  theories with the same  K\"ahler structure but different  complex structures.

%%%%%%%%%%%%%%%%%%%%%%%%%%%%%
  \section{Superconformal  $ {N=(2,2)} $   $\sigma$-models}
\label{sec4}

We turn next  to an algebraic  derivation of the basic formula \eqref{calabi-dist}, 
 which only relies on the  $ N=2$ superconformal  symmetry of the interface. 
 To this end, we  
view an interface  between  two theories,  CFT  and  CFT$^\prime$,  as a formal operator 
mapping the  states on the circle of  CFT  to those of  CFT$^\prime$. This has been  explained for instance
in \cite{Bachas:2001vj,Bachas:2007td}.
Folding converts this  operator   to  a  boundary state
 of  the tensor-product theory $\overline {\rm CFT}\otimes {\rm CFT}^\prime$, where here the bar denotes the parity-conjugate theory.  
  We use the same symbol, $ {\Delta}_f $,  for the operator, 
   for the corresponding brane,  and for its boundary state. 
  Our discussion parallels  the analysis of  $N=2$ superconformal boundaries 
   by Ooguri et al  \cite{Ooguri:1996ck}, and we will  therefore adopt the conventions of these authors.
   \smallskip

Every interface operator  contains  a term\    $g \,  \vert 0 \rangle\langle 0^\prime \vert + \cdots$ ,  where
$\vert 0\rangle$ is the normalized  ground state of theory  CFT,   and $\vert 0^\prime \rangle$  the normalized  ground state
of theory  CFT$^\prime$.  The coefficient of this term is, by definition,   the $g$-function  of the interface. 
Since we will be  working  with non-normalized ground states,    we write more generally
  \be\label{gCyl}
 g \, =\,  {   \langle 0 \vert  q^H \   {\Delta}_f \   q^{H^\prime} \vert 0^\prime \rangle    \over
  \langle 0 \vert  q^H\vert 0\rangle^{1\over 2} \,  \langle 0^\prime \vert  q^{H^\prime} \vert 0^\prime\rangle^{1\over 2}
}\   , 
\ee
where $H$ and $H^\prime$ are the Hamiltonians in the closed-string channel. 
This expression does not depend on the evolution time log\,$q$, so one can  take
 $q\to 0$ and    replace the ground states by any  other states  with non-vanishing vacuum components.

%%%%%%%%%%%%%%

 \subsection{Type-A and type-B boundaries}
 
 We are interested in  interfaces that preserve  a $N=2$ superconformal algebra,   and  
  which can be continuously deformed to the identity operator.
   Since folding converts these operators to boundary states,
  we first recall  some well-known facts  about 
  supersymmetry-preserving boundaries
  in $N=(2,2)$ superconformal theories. 
  
  Supersymmetric branes   come in two varieties, type-A and type-B. 
  The boundary states of type-A  branes obey
  the following conditions:\,\footnote{These are the conditions in the closed-string channel, in which the reality
  conditions for fermions involve the exchange of  left and right movers.
   In the open-string
 channel one has $J_L = -J_R$ and $G_L^+=  G_R^-$ 
   for the type-A boundaries,  $J_L = J_R$ and $G_L^+=  G_R^+$ for the type-B boundaries.}
   \be 
 (G_L^+  -i  G_R^-)\, \vert  A \rangle\hskip -1mm \rangle = (G_L^-  -i  G_R^+)\, \vert  A \rangle\hskip -1mm \rangle =
 (J_{ L} -  J_{ R}) \, \vert  A \rangle\hskip -1mm \rangle    = 0\ , \nonumber 
 \ee 
  \be\label{AA} 
 {\rm and}\qquad e^{i\alpha\phi} \, \vert  A \rangle\hskip -1mm \rangle =
 e^{i\alpha\phi_0}\, \vert  A \rangle\hskip -1mm \rangle\ ,   
 \ee 
where $G^\pm_L $ and $G^\pm_R $
  are the complex left- and right-moving supercurrents, 
$J_L$ and $J_R$ are the R-symmetry currents, and 
 $\phi = \int (J_L - J_R) = \phi_L+\phi_R$.
Likewise, the B-type boundaries obey the conditions
   \be
   (G_L^+  -i  G_R^+)\, \vert  B \rangle\hskip -1mm \rangle = (G_L^-  -i  G_R^-)\, \vert  B \rangle\hskip -1mm \rangle =
 (J_{ L} +  J_{ R}) \, \vert  B \rangle\hskip -1mm \rangle    = 0\ , \nonumber  \ee
   \be\label{BB} 
 {\rm and}\qquad e^{i\alpha \tilde \phi} \, \vert  B \rangle\hskip -1mm \rangle =
 e^{i\alpha \tilde \phi_0}\, \vert  B \rangle\hskip -1mm \rangle\ ,   
 \ee 
 where $\tilde\phi = \int (J_L + J_R) = \phi_L - \phi_R$.
 The above  conditions   
  imply   that type-A branes couple only to the $(c,c)$ and $(a,a)$ fields,  while type-B branes couple to
$(c,a)$ and $(a,c)$ fields. Here  $c$  and  $a$  denote chiral and antichiral primaries of the $N=2$ superconformal algebra,
$(c,c)$ is a field that is chiral with respect to both the left and the right algebra etc.
\smallskip

Another consequence of the above conditions, which will be important for our purposes  here, has to do with spectral flow. 
 The two spectral-flow operators are
  $e^{ic\phi/6}$ and  $e^{ic\tilde\phi/6}$,   with $c$   the central charge of the CFT. 
  It follows then from  \eqref{AA} and \eqref{BB} that
       \be\label{bps}
   \langle\hskip -1mm \langle A  \vert 0 \rangle   =    e^{- i c\phi_0/6}\,  
    \langle\hskip -1mm \langle A  \vert 0 \rangle_{\rm\footnotesize  RR}
\qquad {\rm and} \qquad
 \langle\hskip -1mm \langle B  \vert \tilde 0 \rangle   =    e^{- ic \tilde \phi_0/6 }\,  
    \langle\hskip -1mm \langle B  \vert \tilde 0 \rangle_{\rm\footnotesize  RR}   \ , 
\ee
where  
  $\vert 0\rangle_{\rm\footnotesize  RR}$ and $\vert \widetilde 0\rangle_{\rm\footnotesize  RR}$ are the canonical Ramond-Ramond 
  ground states,  obtained
 from the Neveu-Schwarz vacuum  by   spectral flow.\,\footnote{Readers not familiar with $N=2$ should 
compare these facts to the analogous statements for
  boundary states of a free-boson theory. There, the Dirichlet condition 
 couples only to the closed-string momentum modes,  and the coupling has equal strength
 for  all states that differ only in momentum. This  is necessary in order to  produce  the
localizing $\delta$-function of the Dirichlet brane. Likewise,  a Neumann condition  couples only to   winding modes, and the coupling
depends on winding number  through  a  phase factor.  In the context of $N=(2,2)$ theories, 
type-A and type-B branes are, respectively Dirichlet and Neumann conditions for the field
 $\phi$, and the two spectral-flow operators inject, respectively, momentum and  winding.}

\smallskip

These statements have a nice geometric interpretation in the large volume limit   \cite{Ooguri:1996ck}. 
The  boundary states $ \vert  A \rangle\hskip -1mm \rangle $ correspond to D-branes wrapping Lagrangian submanifolds,
 $\gamma$, of the Calabi-Yau $n$-fold, while the states $ \vert  B  \rangle\hskip -1mm \rangle $
 correspond to ($p$-dimensional)  holomorphic submanifolds
  ${\tilde \gamma}$.  The overlaps with the NS vacuum are the $g$-factors of the corresponding D-branes, whereas
  the overlaps with the appropriately-normalized,
   canonical RR ground states give the  D-brane charges
  \be\label{eq:brane-charge}
  \langle\hskip -1mm \langle A  \vert 0 \rangle_{\rm\footnotesize  RR}   =   \int_\gamma \Omega  
 \qquad {\rm and} \qquad
  \langle\hskip -1mm \langle B  \vert \tilde 0 \rangle_{\rm\footnotesize  RR} = \int_{\tilde \gamma}
 \,  {1\over p !}
\, \omega^p \ . 
 \ee
 In the context  of string theory compactified on a Calabi-Yau 
 3-fold,   the equations \eqref{bps} are  the BPS mass formulae  for the corresponding supersymmetric black holes.

The above disk amplitudes have also an interpretation in terms of  topological twists. To compute  the first   amplitude for example,
 one puts A-type boundary conditions at the end of a semi-infinite cigar, where the curved region at the tip of the cigar is B-twisted. Due to the topological twist, the identity operator sitting at the end point of the cigar becomes a RR-ground state at the end of the cigar. This is the canonical RR ground state, corresponding to the holomorphic 3-form on a Calabi-Yau manifold.

%%%%%%%%%%%%%%%%%%%%%%%%%%%% 
 
  \subsection{Diastasis as entropy of  A-type interfaces}
  
  The discussion of supersymmetric boundaries
  can be adapted easily to N=2 superconformal interfaces \cite{Brunner:2007qu}. 
  These  are of  A-type  or of B-type  
  depending on which  combination of superconformal generators   they intertwine. 
  Explicitly, in terms of the interface operators one has 
   \bea\label{intertwine}
  [ G_L^+  -i  G_R^- ,  \, {\Delta}_{f,A} ] =  
 [  G_L^-  -i    G_R^+ ,  \, {\Delta}_{f,A} ] = [J_{ L} -  J_{ R} , \, {\Delta}_{f,A}] 
 = 0\ , \nonumber 
 \eea
 \bea
  [   G^+_L  -i   G^+_R ,  \, {\Delta}_{f,B} ] =   
  [  G_L^-  -i    G_R^- ,  \, {\Delta}_{f,B} ] =  [J_{ L} +  J_{ R} , \, {\Delta}_{f,B}] = 0\ .  
 \eea
  Likewise,     the intertwining of  spectral-flow operators reads
 \be\label{47}
 e^{i\alpha\phi} \, {\Delta}_{f,A}\, e^{-i\alpha\phi}  = e^{i\alpha\phi_0} \, {\Delta}_{f,A}  \ , \quad {\rm or}\qquad 
 e^{i\alpha\tilde\phi} \, {\Delta}_{f,B}\, e^{-i\alpha\tilde\phi}  = e^{i\alpha\tilde\phi_0} \, {\Delta}_{f,B}  \ , 
 \ee
  where $\phi$  and
  $\tilde\phi$ are the fields defined earlier, and $\phi_0$, $\tilde\phi_0$ are constant phases. 
  These 
  equations are the unfolded versions of equations  \eqref{AA} and \eqref{BB}  for the tensor-product theory.  
  To be precise,   since folding converts the interfaces to boundaries of the product theory
  $\overline {\rm CFT}\otimes {\rm CFT}^\prime$,  the generators that enter in \eqref{AA} and \eqref{BB} 
   are the sums of the generators
  for the individual theories, but with theory CFT parity-transformed.\,\footnote{The tensor-product theory has also
  superconformal branes that do not involve this parity operation. From the geometric point of view, this is because the target  
  manifold is more special than Calabi-Yau. The corresponding interfaces are not continuously-connected
  to the identity, and will not concern us here. } 
    
\smallskip

Combining equations \eqref{gCyl} and  \eqref{47} leads to 
 the following alternative expression for the square of the $g$ function of type-A interfaces:
   \be\label{gCylR}
 g_A^2 \, =\,  {  \ _{\rm\footnotesize  RR}\langle \bar 0 \vert  q^H \,   {\Delta}_{f,A}
  \,   q^{H^\prime} \vert 0^\prime \rangle_{\rm\footnotesize  RR}    \times
  \ _{\rm\footnotesize  RR}\langle \bar 0^\prime \vert  q^{H^\prime} \,  {\Delta}_{f,A}^\dagger
  \,  q^{H} \vert 0  \rangle_{\rm\footnotesize  RR}  
   \over
   \hskip -0.6mm  \ _{\rm\footnotesize  RR}\langle \bar 0 \vert  q^H\vert 0\rangle_{\rm\footnotesize  RR}  \times
 \ _{\rm\footnotesize  RR}\langle   \bar 0^\prime \vert  q^{H^\prime} \vert 0^\prime\rangle_{\rm\footnotesize  RR} 
}\ \  , 
\ee 
 where $\vert   0 \rangle_{\rm\footnotesize  RR}$  is the canonical Ramond-Ramond  ground state obtained  by acting
 with the spectral-flow operator $e^{ic\phi/6}$ on  the Neveu-Schwarz vacuum, 
 and   $\vert \bar 0\rangle_{\rm\footnotesize  RR}$ is the conjugate ground state. 
There is a
 similar expression for type-B interfaces,  with $\vert 0 \rangle_{\rm\footnotesize  RR} $  replaced by
the twisted canonical Ramond ground state,  $\vert \widetilde 0 \rangle_{\rm\footnotesize  RR} $, obtained 
with the spectral-flow operator $e^{ic\tilde \phi/6}$. 
 Taking the logarithm of   \eqref{gCylR}  
gives an expression  reminiscent  of the  diastasis function  of the previous section, 
\be\label{410}
2\, {\rm log} (g_{A}) \,  =  K(t, \bar t)  + K(t^\prime, \bar t^\prime)   - 
{\rm log} ( \hskip -0.6mm \ _{\rm\footnotesize  RR}\langle \bar 0 \vert     {\Delta}_{f,A}
    \vert 0^\prime \rangle_{\rm\footnotesize  RR})  -
    {\rm log} ( \hskip -0.6mm \ _{\rm\footnotesize  RR}\langle \bar 0^\prime \vert     {\Delta}_{f,A}^\dagger
    \vert 0  \rangle_{\rm\footnotesize  RR} ) \ . 
\ee
We used here the fact  \cite{Cecotti:1991me} (see also \cite{Periwal:1989mx})
that  the   canonical Ramond ground state, which  has holomorphic dependence
   on the complex structure moduli, has  norm  
    \be
 {\rm log}  \left(\hskip -0.4mm  \ _{\rm\footnotesize  RR}\langle   \bar 0    \vert 0 \rangle_{\rm\footnotesize  RR} \right)
 \ =\    - K(t, \bar t) \ . 
   \ee
 There are analogous expressions for type-B interfaces  with$K(t, \bar t)$  replaced by 
$K(u, \bar u)$, the K\"ahler potential on the moduli space of K\"ahler structures. 

\smallskip

To show that \eqref{410}   is Calabi's diastasis we interpret the expression 
 ${\rm log} ( \hskip -0.6mm \ _{\rm\footnotesize  RR}\langle \bar 0 \vert     {\Delta}_{f,A}
    \vert 0^\prime \rangle_{\rm\footnotesize  RR} ) $ at large volume. For this, we note that $ \vert 0  \rangle_{\rm\footnotesize  RR} $  becomes the holomorphic three-form for the geometry 
    corresponding to the unprimed theory, whereas $\hskip -0.6mm \ _{\rm\footnotesize  RR}\langle \bar 0 \vert$ corresponds to the anti-holomorphic three-form for the primed  theory. The expression then becomes,  in the folded picture,  the RR-charge of an A-type brane with respect to the canonical RR ground state. The relevant A-brane is the deformed diagonal brane described before. Hence, we can employ (\ref{eq:brane-charge}) to conclude that at large volume
\begin{equation}\label{eq:analytic}
\hskip -0.6mm \ _{\rm\footnotesize  RR}\langle \bar 0 \vert     {\Delta}_{f,A}
    \vert 0^\prime \rangle_{\rm\footnotesize  RR} = \int_M \bar\Omega' \wedge f^* \Omega \ .
\end{equation}
This is precisely the analytic continuation of the K\"ahler potential appearing in Calabi's diastasis function. 
Note that relation (\ref{eq:analytic}) was computed at large volume. However, the relevant disk one-point functions  do not 
depend on  K\"ahler moduli, hence the above relation can be extrapolated to all length scales.

  \subsection{Quantum diastasis function}  
 
 We will now prove that  for any B-twistable theory the left hand side of (\ref{eq:analytic})  depends holomorphically on  $t^\prime$ 
 and   antiholomorphically on  $t$, in some open region of moduli space.
 The proof does not rely on a  geometrical interpretation, and  also goes through practically unchanged for B-type interfaces
 in A-twistable theories. This shows that
 \bea\label{analytic43}
 \hskip -0.6mm \ _{\rm\footnotesize  RR}\langle \bar 0 \vert     {\Delta}_{f,A}
    \vert 0^\prime \rangle_{\rm\footnotesize  RR} \equiv  e^{-K(t^\prime, \bar t)}\ , \quad {\rm and}
    \quad 
     \hskip -0.6mm \ _{\rm\footnotesize  RR}\langle \bar {\tilde 0} \vert     {\Delta}_{f,B}
    \vert {\tilde 0}^\prime \rangle_{\rm\footnotesize  RR} \equiv  e^{-K(u^\prime, \bar u)}
 \eea
 define  the analytic extensions of the quantum K\"ahler potentials to independent holomorphic and anti-holomorphic moduli.
 \smallskip
 
To establish this holomorphicity property, recall that N=2 
supersymmetric conformal   theories can be marginally perturbed by suitable descendants of fields from the chiral $(c,c)$ sector, or the twisted chiral $(a,c)$ sector. The two are interchanged by mirror symmetry, and we  focus here  on chiral perturbations, which geometrically correspond to complex structure deformations. 
 On the level of the action, and on a worldsheet with boundaries,   the perturbation reads
\begin{equation}
S \to S + \Delta S (t_i) + \overline{\Delta S}(\bar{t}_i),
\end{equation}
where 
\begin{equation}
\Delta S = \sum_i t_i \int Q_L^+ Q_R^+ \phi_i + \Delta S_{b}\ ,    \qquad
\overline{\Delta S} = \sum_i \bar{t}_i \int Q_R^- Q_L^- \bar\phi_i + \overline {\Delta S_{b}} \ , 
\end{equation}
and 
\begin{equation}\label{Sbound}
\Delta S_{b} + \overline {\Delta S_{b}}\ =\  -2i \oint_{\cal C} ds  (t_i \phi_i-\bar{t}_i\bar\phi_i) \ .
\end{equation}
 The addition of this boundary term is necessary 
 in order to preserve   A-type supersymmetry  \cite{Hori:2000ck},  assuming this
 was the unbroken supersymmetry in the unperturbed theory. 
The supercharges $Q_L^\pm, Q_R^\pm$ are obtained by contour integration of the 
supersymmetry  currents $G_L^\pm, G_R^\pm$, and they  satisfy the standard $N=2$ algebra
\begin{equation}
\{ Q_L^+, Q_L^- \} = 2 (H+P), \quad \{ Q_R^+, Q_R^- \} = 2 (H-P) \ ,
\end{equation}
where $H$ and $P$ are translation operators in Euclidean time and space.

\begin{figure}[tbp]
\centering 
\includegraphics[width=.75\textwidth,trim=0cm 8cm 0cm 6cm,clip=true]{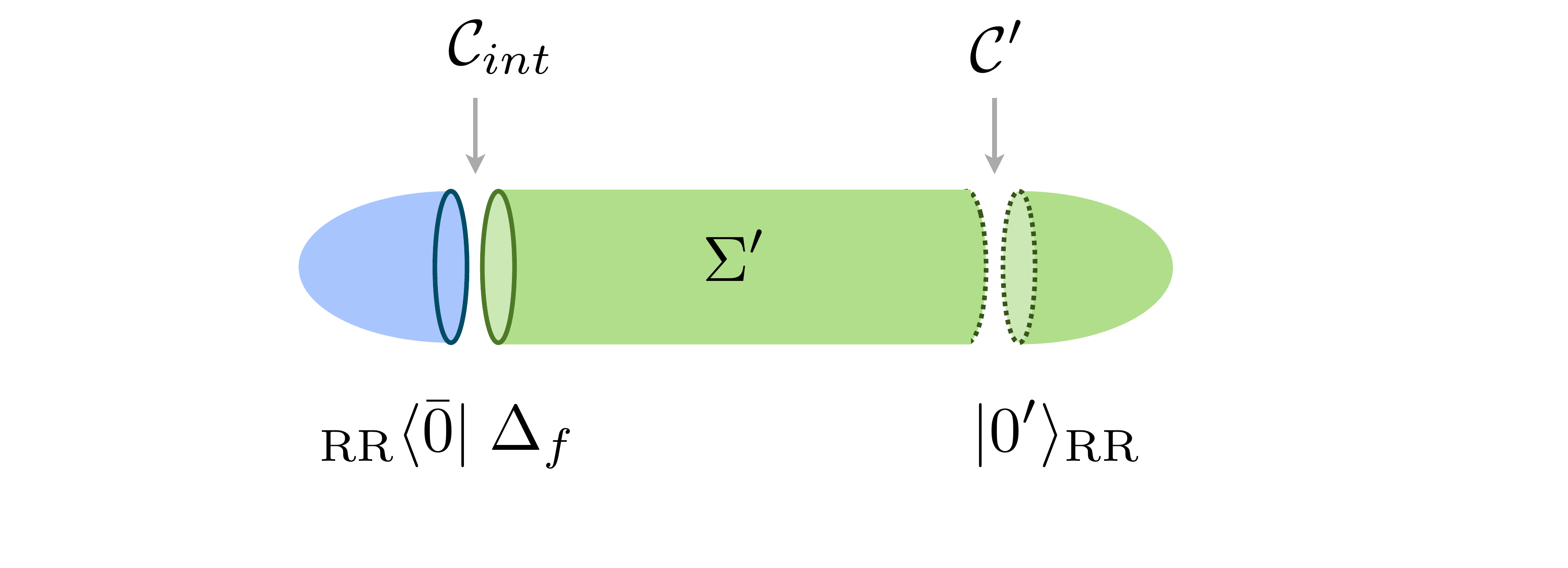}
\caption{\label{fig:i} Schematic representation of the amplitude \eqref{analytic43} that defines the
analytic extension of the K\"ahler potential to independent holomorphic and antiholomorphic moduli.}
\end{figure}

We are interested in  perturbations that are restricted to only part  of the worldsheet. The resulting interface then connects an unperturbed initial theory to another theory on the same moduli space.  If the perturbation is from the $(c,c)$   sector,  the   interface will be A-type,  i.e. 
 it preserves the same supersymmetry as A-type branes.
Thus the fusion between $(c,c)$ deformation interfaces and A-type branes is protected by supersymmetry, in agreement with the fact that A-type branes remain supersymmetric under $(c,c)$   perturbations \cite{Brunner:2007ur}. 
Viewed as an operator, the A-interface acts naturally on elements of the  $(c,c)$   ring, in the same way that A-branes couple to  $(c,c)$ fields.
 \smallskip
 
   Recall from the discussion in section \ref{sec3} that complex structure interfaces were related 
   after folding to special Lagrangian branes. The deformation of the original diagonal brane was determined
   by  a map  $f: M\to M^\prime$,  which ensured  that the deformed brane is still  special Lagrangian, hence supersymmetric.
 In the conformal theory, the role of $f$ is played by the boundary perturbation (\ref{Sbound}), which  adjusts  to 
 the bulk perturbation continuously as long as there are no relevant operators at the interface. After a 
 finite perturbation, on the other hand,  the O.P.E. of the perturbing field with the boundary may stop being regular,
 thereby   inducing a renormalization-group  flow to some lower-$g$ interface.  When   a space-time interpretation of D-branes
  in terms of BPS states is available, this means that we hit a line of marginal stability.
\smallskip

  From now on, we assume that relevant boundary operators do not appear, which should be true  in open regions around
  generic points of moduli space. We would like to show that 
 \begin{equation}\label{eq:hol-derivative}
\frac{\partial}{\partial \bar{t_i}^\prime  } \,  \left[ {}_{\rm\footnotesize RR}\langle \bar 0 | \Delta_f | 0^\prime  \rangle_{\rm\footnotesize RR}\right]   = 0 \ ,
\end{equation}
  so that the amplitude has holomorphic dependence on the primed moduli.   
    Our analysis follows \cite{Hori:2000ck}. 
The amplitude under consideration is drawn  in Figure 1. We model the 
      RR ground state $|0'\rangle_{\rm\footnotesize RR} $ by a semi-infinite cigar in a B-twisted topological theory. 
       Since the operator insertion at the tip of the cigar is the identity, the state appearing at the boundary
       of the cigar  is the canonical RR ground state. 
       Similarly, we create the state ${}_{\rm\footnotesize RR} \langle \bar 0 |$ by inserting the identity at the tip of a semi-infinite cigar with 
       an anti-topological B-twist.  The two cigars 
are connected by a   flat cylinder, on which we locate the interface $\Delta_f$ that separates  the perturbed
  (green)  region from  the unperturbed (blue) region.

\smallskip 
  
  We can now  prove  \eqref{eq:hol-derivative} in two steps. First, we use   the results of \cite{Cecotti:1991me} to conclude that
the canonical RR ground state has holomorphic dependence on the moduli,  
\begin{equation}
\frac{\partial}{\partial \bar{t}_i^\prime} \, |0'\rangle_{\rm\footnotesize RR}\ =\ 0 \ . 
\end{equation}
 Hence, all we need to do is to consider the $t_i$ perturbation on a flat region $\Sigma'$  
 between  the interface and the boundary of the semi-infinite cigar, see Figure 1. 
   Taking the derivative in (\ref{eq:hol-derivative}) then amounts to inserting  in the amplitude
\begin{equation} 
\overline {\Delta S_i} = \int_{\Sigma'} Q_R^- Q_L^- \bar{\phi}_i +2i \oint_{{\cal C}_{int}} \bar{\phi}_i \ . 
\end{equation}
 Using the supersymmetry algebra, we can rewrite this insertion as 
 \begin{equation}\label{insert}
\overline {\Delta S_i} = \int_{\Sigma'} (Q_R^- - i Q_L^+)(Q_L^-+i Q_R^+) \bar\phi_i  + 2i \oint_{{\cal C}'} \bar{\phi}_i \ ,
\end{equation}
 where ${\cal C}'$ is the boundary of the cigar on the right. The above rewriting   follows  from the 
 supersymmetry algebra and the fact that $\bar \phi_i$ is a anti-chiral field. Together these imply that
 \bea
  (Q_R^- - i Q_L^+)(Q_L^-+i Q_R^+) \bar\phi_i = (Q_R^- - i Q_L^+)Q_L^- \bar\phi_i  
=  
  (Q_R^-Q_L^- -2 i H) \bar\phi_i\ ,   
\eea 
 where $H$ is the generator of translations perpendicular to the interface. 
 \smallskip

Proving equation  \eqref{eq:hol-derivative}  is  therefore equivalent to proving that
  \bea {}_{\rm\footnotesize RR}\langle \bar 0 | \Delta_f \, \left[ \int_{\Sigma'} (Q_R^- - i Q_L^+)(Q_L^-+i Q_R^+) \bar\phi_i  + 2i \oint_{{\cal C}'} \bar{\phi}_i 
  \right] \,  | 0^\prime  \rangle_{\rm\footnotesize RR}   = 0\ . 
\eea
    The second piece vanishes,   since  otherwise the amplitude  would violate the  selection rules for R-charge. 
As for the bulk  piece, 
we  use that $Q_R^--i Q_L^+$ intertwines the interface operator $\Delta_f$. 
 By contour deformation, we can thus let it act on the bra  state ${}_{\rm\footnotesize RR} \langle \bar{0}|$, 
 which is annihilated  since it is  a ground state. Here, we have used that $\Sigma'$ is flat, hence all contours can be deformed freely. 
This concludes the proof of (\ref{eq:hol-derivative}).

  %%%%%%%%%%%%%%%%%%
  
\section{On the triangle inequality}
\label{sec5}
   
 We conclude this letter with a remark on whether Calabi's diastasis function defines a  distance. 
 Although Calabi did not comment on this in \cite{Calabi},
the name he chose suggests that he knew it did not.   
Since his  motivation   was to study isometric embeddings of K\"ahler manifolds, this question was
not central to his work anyway. If, on the other hand,  
   the proposal \eqref{distance} for a distance between
 conformal field theories \cite{inprogress} makes sense, it should do so  for
 Calabi-Yau manifolds, for which we have proven  
   equation \eqref{calabi-dist}. The question, therefore, is whether
     the square-root  of the diastasis function defines  a distance in the special case of  
 Calabi-Yau moduli spaces.  
 
 \smallskip 
 
  Actually, the complex structure moduli space of a Calabi-Yau $n$-fold can be embedded
isometrically  in a higher-dimensional 
projective space. This is because the K\"ahler potential 
 $ K = -\log (\int_M \Omega\wedge \bar \Omega)$ 
  can be interpreted as the
 Fubini-Study potential 
\be \label{eq:fs-indef}
K = -\log \eta_{\alpha\beta} \Pi^\alpha \bar\Pi^\beta 
\ee
restricted to the embedding
\be
t\rightarrow \Pi^\alpha \equiv \int_{\Sigma_\alpha} \Omega \ , 
\ee
where $\Sigma_\alpha$ is a basis for $H_{n}(M,\mathbb{Z})$, and $\eta_{\alpha\beta}$ is the
intersection form (antisymmetric for $n=2k+1$).
 As a first step, we may thus like to 
   check whether the square root of   
Calabi's  diastatic function on projective space is a distance. 
\smallskip
 
  The answer actually depends on the signature of  the metric.  
It is true in the  hyperbolic case  (signature $-+++\ldots$), which  includes the 
complex structure and K\"ahler moduli spaces on $T^2$.
To see why, 
  take a coordinate patch $z_0=1$ and choose, without loss of generality,  the third
point  at $\vec z=0$. If the  other two points are at $\vec z_1$ and $\vec z_2$  
we have
\bea
d_{i3}^2 &=& -\log (1-|\vec z_i|^2) \qquad i=1,2 ;  \nonumber \\
d_{12}^2 &=& \log \frac{|1-\vec z_1\cdot\bar z_2|^2}{(1-|\vec z_1|^2)(1-|\vec z_2|^2)}  
 =  \log |1-\vec z_1\cdot\bar z_2|^2 +  d_{13}^2 + d_{23}^2 .
\eea
To check the triangle inequality, we  need to check that 
\be \label{eq:tri-sq}
(d_{13} + d_{23})^2 \ge d_{12}^2 \Longleftrightarrow 
2 d_{13} d_{23} \ge 2 \log |1-\vec z_1\cdot\bar z_2| .
\ee
 The worst case is when $z_1=x$ and $z_2=-x$ are antialigned, so it is sufficient to check that
 \be
|\log (1-x^2)| \ge \log (1+x^2) . 
\ee
This inequality  is indeed  true for all values of $x$. 
\smallskip 

The same analysis for the elliptic case
(signature $+++\ldots$) shows  that 
the triangle inequality fails.  Take for instance the three points  $z_1=z$, $z_2=1$ and $z_3=0$
on $CP^1$ for which one finds
 \be
d^2_{13} (0, z) = \log (1+ |z|^2) \ , \quad d^2_{12}   = \log \frac{2(1+| z|^2)}{|1+z |^2}\ ,
\quad  d^2_{23}   = \log 2\ . 
 \ee
Since $d_{13}$ 
 diverges as $ z\rightarrow \infty$, while  $d_{12}$ and $d_{23}$ approach $\sqrt{\log 2}$, the triangle inequality
cannot possibly be valid. 
  What about the projective spaces relevant for  
Calabi-Yau  moduli spaces?  For general signature $(n,m)$ this contains
the elliptic case as a submanifold, so it cannot hold everywhere.  Thus the actual
embedding is important.

 While one could study this,
it is simpler to test the claim locally by constructing the expansion for the diastatic function from
the intrinsic geometric data of the moduli space.  In K\"ahler normal coordinates, we
have
\be
K(t,\bar t) = \sum_\alpha |t^\alpha|^2
  - \frac{1}{4} R_{\alpha\bar\beta \gamma\bar\delta} t^\alpha \bt^{\bar\beta} t^\gamma \bt^{\bar\delta}
   + \ldots
\ee
so the Calabi diastasis function is
\be
d^2(t_1,t_2) = |t_1-t_2|^2
  - \frac{1}{4} R_{\alpha\bar\beta \gamma\bar\delta} 
  ( t_1^\alpha t_1^\gamma -  t_2^\alpha t_2^\gamma )
  ( \bt_1^{\bar\beta} \bt_1^{\bar\delta} - \bt_2^{\bar\beta} \bt_2^{\bar\delta} )
   + \ldots
\ee
Setting  $t_3=0$, the triangle inequality at this order becomes
\bea
 |t_1| + |t_2| - |t_1-t_2|   && \geq  -  \frac{1}{8} R_{\alpha\bar\beta \gamma\bar\delta} 
  \Biggl[  \frac{1}{|t_1-t_2|} 
  ( t_1^\alpha t_1^\gamma -  t_2^\alpha t_2^\gamma )
  ( \bt_1^{\bar\beta} \bt_1^{\bar\delta} - \bt_2^{\bar\beta} \bt_2^{\bar\delta} ) 
\nonumber \\
 &&
  - \frac{1}{|t_1|}  t_1^\alpha t_1^\gamma \bt_1^{\bar\beta} \bt_1^{\bar\delta}
  - \frac{1}{|t_2|}  t_2^\alpha t_2^\gamma \bt_2^{\bar\beta} \bt_2^{\bar\delta}
  \Biggr]\,  .
\eea
For the special choice  $t_1=x$ and $t_2=-x$, one finds $R_{x\bx x\bx} <0$, namely the 
sectional curvature in the plane $x\bx$ must be negative. This confirms 
   the previous computation, and shows that
there is a local condition.  

\smallskip
Since K\"ahler manifolds can have either sign of sectional curvature, we see that
in general the diastatic function does not satisfy the triangle inequality.  For a Calabi-Yau moduli
space, in particular, the sectional curvature is known to be positive  near a conifold point \cite{Candelas:1990rm}. 
 Thus, if 
 the supersymmetric interfaces minimize  the entropy, 
 the proposal \eqref{distance} violates  the triangle inequality. We hope to return to  this problem 
 in the  future  \cite{inprogress}.

    %%%%%%%%%%%%%%%%%

\acknowledgments

The work of I.B. is supported in part 
by a EURYI grant of the European Science Foundation.  The work of M.D. is supported in part by DOE grant DE-FG02-92ER40697.
The work of
L.R. is supported in part by the NSF under Grant PHY-0969919 and by a Simons Foundation Sabbatical Fellowship.

We have benefited from useful discussions with Amir Kashani-Poor, Michael 
Kay, Bruno Le Floch and Vasilis Niarchos. C.B. thanks the string-theory group at LMU for their warm hospitality
during an extended visit made possible by the Humboldt foundation.  L.R. is grateful to the theory groups of the LPTENS, Paris,
and of the IAS, Princeton,  for their warm hospitality.

\end{document}